%% file: main_is2019.tex
\documentclass[a4paper]{article}

\usepackage{INTERSPEECH2019}

\usepackage{amsmath,graphicx}
\usepackage{cite}
\usepackage{url}
\usepackage{multirow}
\usepackage{subfig}
\usepackage{microtype}
\usepackage{mathrsfs}
\usepackage{xcolor}
\usepackage{booktabs}
\usepackage{amssymb}
\usepackage{calc}
\usepackage[prependcaption,textsize=scriptsize,disable]{todonotes}

\definecolor{mycolor}{HTML}{FF6600}

\title{Feature exploration for almost zero-resource ASR-free keyword spotting using a multilingual bottleneck extractor and correspondence autoencoders}

\name{Raghav Menon$^1$, Herman Kamper$^1$, Ewald van der Westhuizen$^1$, John Quinn$^{2,3,4}$, Thomas Niesler$^1$}
\address{
$^1$Department of Electrical and Electronic Engineering, Stellenbosch University, South Africa\\
$^2$UN Global Pulse, Kampala, Uganda \\
$^3$Department of Computer Science, Makerere University, Uganda\\
$^4$School of Informatics, University of Edinburgh, UK
}

\email{rmenon@sun.ac.za kamperh@sun.ac.za, ewaldvdw@sun.ac.za, trn@sun.ac.za}

\begin{document}

\maketitle


\begin{abstract}
We compare features for dynamic time warping (DTW) when used to bootstrap keyword spotting (KWS) in an almost zero-resource setting.
Such quickly-deployable systems aim to support United Nations (UN) humanitarian relief efforts in parts of Africa with severely under-resourced languages.
Our objective is to identify acoustic features that provide acceptable KWS performance in such environments.
As supervised resource, we restrict ourselves to a small, easily acquired and independently compiled set of isolated keywords.
For feature extraction, a multilingual bottleneck feature (BNF) extractor, trained on well-resourced out-of-domain languages, is integrated with a correspondence autoencoder (CAE) trained on extremely sparse in-domain data.
On their own, BNFs and CAE features are shown to achieve a more than 2\% absolute performance improvement over baseline MFCCs.
However, by using BNFs as input to the CAE, even better performance is achieved, with a more than 11\% absolute improvement in ROC AUC over MFCCs and more than twice as many top-10 retrievals for two evaluated languages, English and Luganda.
We conclude that integrating BNFs with the CAE allows both large out-of-domain and sparse in-domain resources to be exploited for improved ASR-free keyword spotting.
\end{abstract}

\section{Introduction}
\label{sec:intro}

In Uganda, internet infrastructure is often poorly developed, precluding the use of social media to gauge sentiment.
Instead, community radio phone-in talk shows are used to voice views and concerns.
In a project piloted by the United Nations (UN), radio browsing systems have been developed to monitor such radio shows~\cite{Menon2017,Saeb2017}.
Currently, these systems are actively and successfully supporting relief and developmental programmes by the organisation. 
However, the deployed radio browsing systems use automatic speech recognition (ASR) and are therefore highly dependent on the availability of substantial transcribed speech corpora in the target language. 
This has proved to be a serious impediment when quick intervention is required, since the development of such a corpus is always time-consuming.

In a conventional keyword spotting system, where a speech database is searched for a set of keywords, ASR 
is used to generate lattices which are in turn searched for the presence or absence of keywords~\cite{Larson12, mandal2014recent}.
In resource-constrained settings where ASR is not available and cannot be developed, ASR-free keyword spotting approaches become attractive, because these are developed without substantial labelled data~\cite{audhkhasi+etal_icassp17, Menon_Inter2018, Menon_SLTU2018, SainathPara2015, chen2014small, tang2018deep}.
\todo[color=green]{RM: Maybe we will have to give some reason on why we chose DTW over segmental DTW in introduction}
\todo{Ewald: Herman, can you help with this suggestion from Raghav?}
One approach to ASR-free keyword spotting is to extend query-by-example search (QbE), where the search query is provided as audio rather than a written keyword.
\todo[color=gray]{HK: Our DTW approach can be seen as a simplified version of segmental DTW (it isn't straight-forward DTW). But I wouldn't talk about this in the intro. I added one more phrase in Section 4.}
QbE can be performed by using dynamic time warping (DTW) to perform a direct match between a search query and utterances in the search collection~\cite{Hazen2009, Zhang2009,park+glass_taslp08,jansen+vandurme_interspeech12}. 
This approach uses a number of labelled spoken keyword instances as templates.
Each template is used as a query for the DTW-based QbE.
Since the class of each template is known, the individual per-exemplar QbE results can be aggregated to determine whether a certain keyword occurs in a particular utterance.
The advantage of this approach is that only a small set of labelled keywords is required and not a large transcribed corpus as used for ASR-based keyword spotting~\cite{Menon_Inter2018, Menon_SLTU2018}.

Recent interest in zero-resource QbE has led researchers to consider the use of various features~\cite{vavrek2012tuke,carlin2011rapid,jansen2012jhu,lopez2016finding, dunbar2017zero,versteegh2016zero,renshaw+etal_interspeech15}.
Among these, multilingual bottleneck feature (BNF) extractors, trained on well-resourced but out-of-domain languages, have been shown to improve on the performance of MFCCs~\cite{Menon_SLTU2018,Hermann2018,hermann+etal_submitted18,vesely2012language,vu2012investigation, cui2015multilingual, alumae2016improved,sthomas2012,chen2017multilingual, yuan2017extracting}.

Our goal is to improve DTW-based keyword spotting by combining the advantages of using labelled resources from well-resourced languages for learning features, with the advantage of fine-tuning on extremely sparse labelled data in the low-resource target language.
For fine-tuning on target data, we use the correspondence autoencoder (CAE), a model originally developed for the zero-resource setting where only unlabelled data is available \cite{kamper2015unsupervised,renshaw+etal_interspeech15}.
As target language data, we use a small number of labelled isolated keywords that can be easily and quickly gathered.
These keyword instances do not form part of the radio talk show training and evaluation data and can thus be considered out-of-corpus augmentation data.
By learning a mapping between all possible combinations of alternative utterances of the same keyword type, the CAE can learn to disregard aspects not common to the keywords, such as speaker, gender and channel, while capturing aspects that are, such as word identity.
Our work builds on the ideas established in \cite{Hermann2018,hermann+etal_submitted18}, where a CAE trained on BNFs using a large set of in-corpus, ground truth word pairs outperformed other methods in intrinsic evaluations.
This improvement, however, did not hold consistently when automatically discovered word segments were used, in which case the CAE training was completely unsupervised.
\todo{trn: Do you mean "... CAE training was completely unsupervised"?}
\todo[color=green]{RM: CAE as such is an autoencoder which is trained on word alignments obtained from DTW hence completely unsupervised. when using CAE, the only known data is the keywords. Our case cannot be totally unsupervised as we use BNFs as well. Please note that the DTW used here to get the alignments is totally independent of the DTW used for keyword spotting}
In
\todo{Ewald: I'm not sure if ``In contrast'' is entirely apt, but I'm trying to differentiate this work from the others.}
contrast, we show here that consistent improvements can be obtained by combining BNFs with a CAE when fine-tuning on a small number of out-of-corpus gathered keyword instances, i.e.\ lightly supervised.
\todo{Would you classify our work also as 'completely unsupervised' (I don't think so). Perhaps 'lightly supervised'? Then, perhaps, "In contrast, we show here that, when lightly supervised using the small out-of-corpus dataset, consistent ..."}

We benchmark CAE features against MFCCs and BNFs and show that, when a CAE is trained on top of the BNFs, best keyword spotting results are achieved.
This indicates that multilingual feature extraction and target language fine-tuning can be complementary.
We evaluate our approach for two languages: English, which is a proxy language for experimentation; and Luganda, which is a low-resource language of current interest for humanitarian relief efforts.

\section{Radio browsing system}
\label{sec:rbs}

\begin{figure}[t]
	\centering
	\includegraphics[width=.95\linewidth]{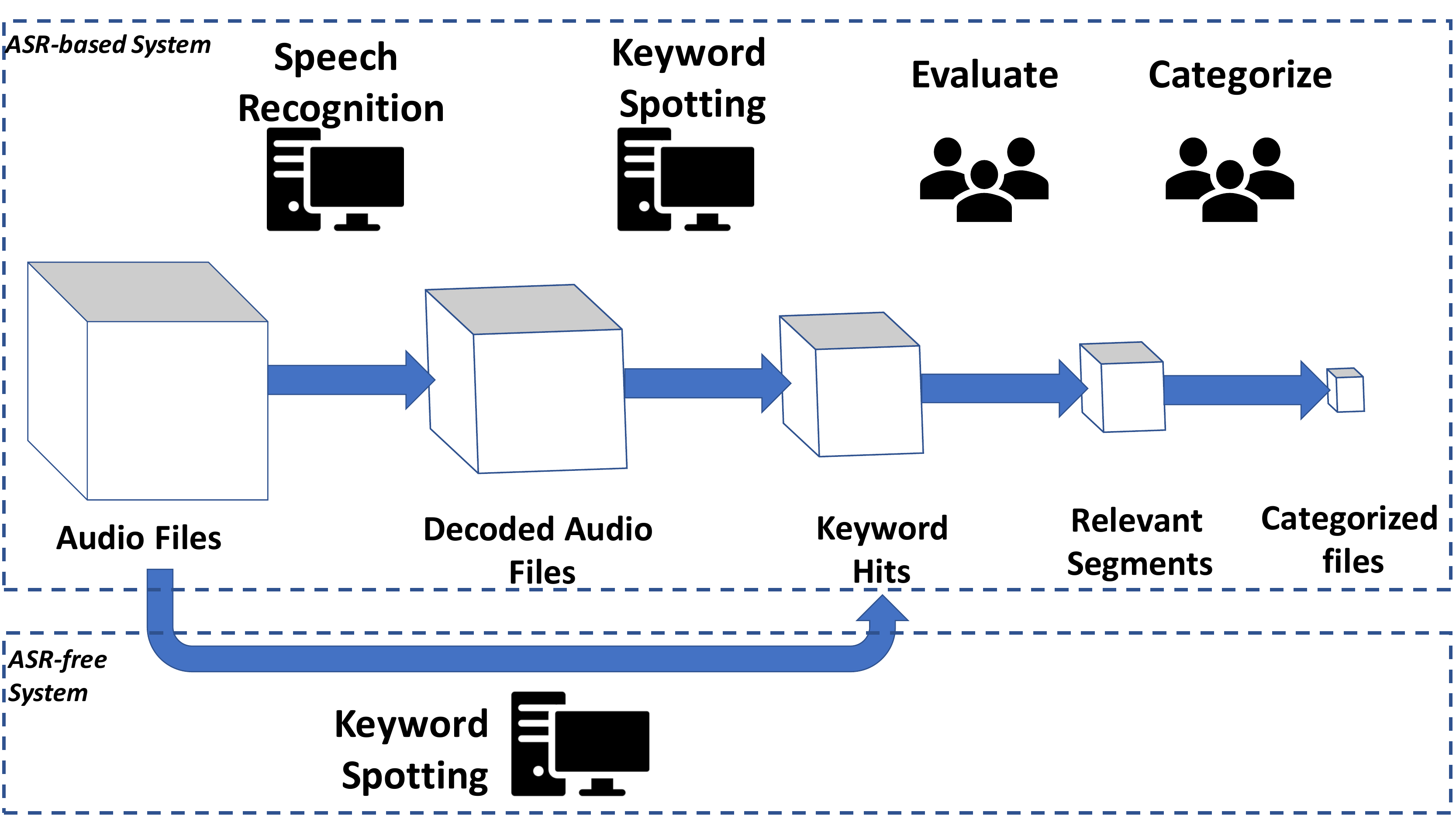}
	\caption{The United Nations (UN) radio browsing system.}
	\label{fig:rbs}
\end{figure}

The existing UN radio browsing system, shown in the top half of Figure~\ref{fig:rbs}, uses ASR to decode the audio and produces lattices that are searched for keywords.
Human analysts filter the detected keywords and their metadata is compiled into a structured, categorised and searchable format.
The ASR-free system (bottom half) bypasses the ASR and lattice search by detecting occurrences of the keywords directly in the incoming audio~\cite{Menon_Inter2018, Menon_SLTU2018}.
High false positive keyword spotting rates can be accommodated due to the presence of {the} human analysts, and the output of the system as a whole has been in continuous successful operation for several months.
A more detailed discussion on the role of human analysts and the detected topics of interest has been presented in~\cite{Saeb2017}.\footnote{Examples available at~\mbox{\url{http://radio.unglobalpulse.net}.}}

\section{Data}
\label{sec:data}
We used a 23-hour English corpus of South African Broadcast News (SABN)~\cite{Kamper2015} and a 9.6-hour corpus of Luganda phone-in talk radio speech as search data in two separate experiments.
Since transcriptions are available for these data sets, it allows system performance to be experimentally evaluated.
However, in all other respects we consider the data as untranscribed.
English is used as a proxy on which we can perform extensive evaluation, while the implementation in Luganda is a practical application of the system in a truly low-resource language.
Table~\ref{tab:SABC_data} shows how the corpora have been split into training, development and test sets.

To train the English keyword spotter, we use a small independent corpus of 40 isolated keywords, each uttered at least once by 24 South African speakers (12 male, 12 female).
The resulting set of 1160 isolated keyword utterances represents the only labelled in-domain data the English keyword spotter uses for training.
There is no speaker overlap with the SABN dataset, which is treated exclusively as search data.

To train the Luganda keyword spotter, we use a small independent corpus of 18 isolated keywords uttered by various male and female speakers in varying recording conditions.
Approximately 32 utterances per keyword type were retained after performing quality control on the recordings.
The resulting set of 603 isolated keyword utterances represents the only labelled in-domain data our keyword spotter uses for training.
There is no speaker overlap with the Luganda talk radio dataset, which is treated exclusively as search data.
\todo{Ewald: Let me know if we should drop the Luganda dev set results altogether, then we can rephrase the following sentences relating to the dev and test sets.}
\todo[color=green]{i do not think it is a good idea to drop the dev set results as all the results are discussed based on dev set.}
Seven keyword types which had frequencies higher than 10 in the corpus development were retained for evaluation against the development set.
This was done to avoid errors in calculating the metrics caused by very low and zero frequency keywords.
For the test set, the full set of keywords was used for evaluation.

The mismatch between the query and search datasets for both languages is intentional as it reflects the operational setting of the radio browsing systems.


\section{Dynamic time warping-based \\ keyword spotting}
\label{sec:DTW}
Dynamic time warping (DTW) is an appropriate approach to keyword detection when only a few isolated exemplars of keywords are available, because it requires as little as a single audio template. 
DTW aligns two time series, represented as feature vector sequences, by warping the relative time axes iteratively until an optimal match is found. 

\begin{table}[!t]
\centering
\caption{
The South African English Broadcast News (SABN) and Luganda datasets.
(\#utts: Number of utterances;
dur: Speech duration in hours;
Dev: Development set.)
}
\vspace*{-5pt}
\label{tab:SABC_data}
\renewcommand{\arraystretch}{1.2}
\begin{tabular}{lcccc}
\toprule
\multirow{2}{*}{\textbf{Set}} &              \multicolumn{2}{c}{\textbf{English}}              & \multicolumn{2}{c}{\textbf{Luganda}} \\
\cmidrule(lr){2-3} \cmidrule(l){4-5}
             &        \textbf{\#utts}        &        \textbf{dur}         & \textbf{\#utts} &     \textbf{dur}      \\ \midrule
Train        & \leavevmode\hphantom{0}5\,231 & \leavevmode\hphantom{0}7.94 &     6\,052      &         5.57          \\
Dev          & \leavevmode\hphantom{0}2\,740 & \leavevmode\hphantom{0}5.37 &     1\,786      &         2.04          \\
Test         & \leavevmode\hphantom{0}5\,005 &            10.33            &     1\,420      &         1.99          \\
Total        &            12\,976            &            23.64            &     9\,258      &         9.06          \\ \bottomrule
\end{tabular}
\end{table}




For DTW-based keyword spotting, features are extracted for both the keyword exemplar and the search utterance in which the keyword is to be detected. 
In our straightforward implementation, the keyword exemplar is slid progressively over the search utterance and at each step DTW computes the alignment cost between the keyword and the portion of the utterance under alignment. 
Using a step of 3 frames, the overall best alignment for each search utterance is determined and taken as a score indicating how likely it is that the search utterance contains the keyword.
Since we have more than one exemplar of the same keyword type, the best score across all templates of the same keyword type is used. 
By applying an appropriate threshold to this score, a decision can be taken regarding the presence or absence of the keyword in each search utterance.   
More refined DTW-based search approaches have been proposed~\cite{Hazen2009, Zhang2009,park+glass_taslp08,jansen+vandurme_interspeech12}, {mainly to improve efficiency}, but here we restrict ourselves to this straightforward implementation.
Future work will consider more advanced matching approaches.
\todo[color=gray]{HK: Added `mainly to improve efficiency' and `Future work ...'.}



\section{Neural network feature extraction}
\label{sec:FNN}

We investigate different types of input features for our DTW-based keyword spotter.
While transcribed in-domain data is difficult, time-consuming and expensive to compile, untranscribed in-domain speech {audio} data is much easier to obtain in substantial quantities.
We investigate the use of autoencoders and correspondence autoencoders as a means of taking advantage of such untranscribed data. The latter requires a sparse set of labelled examples in the target language.
In addition, although large amounts of transcribed in-domain speech data may not be available, large annotated speech resources do exist for several well-resourced languages.
These datasets can be used to train multilingual bottleneck feature extractors.

\subsection{Autoencoder features}
\label{sec:AE}
An autoencoder (AE) is a feedforward neural network trained to reconstruct its input at its output. 
A single-layer AE consists of an input layer, a hidden layer and an output layer. 
The AE takes input $\mathbf{x} \in \mathbb{R}^{D}$ and maps it to a hidden representation $\mathbf{h}=\sigma(\mathbf{W}^{(0)}\mathbf{x}+\mathbf{b}^{(0)})$, with $\sigma$ denoting a non-linear activation (we use $\mathrm{tanh}$).
The output of the AE is obtained by decoding the hidden representation: $\mathbf{y}=\sigma(\mathbf{W}^{(1)}\mathbf{h}+\mathbf{b}^{(1)})$.
The network is trained to reconstruct the input using the loss $||\mathbf{x} - \mathbf{y}||^2$.

A stacked AE~\cite{gehring2013extracting} is obtained by stacking several AE{s}, each AE-layer taking as input the encoding from the previous layer.
The stacked network is trained one layer at a time, each layer minimizing the loss of its output with respect to its input.
A number of studies have shown that hidden representations from an intermediate layer in such a stacked AE are useful as features in speech applications~\cite{kamper2015unsupervised,zeiler+etal_icassp13, badino2014auto,hinton2012deep,deng2010binary,sainath2012auto,gehring2013extracting}.

We train an 8-layer stacked AE feature extractor on the training set shown in Table~\ref{tab:SABC_data}, disregarding the transcriptions.
39-dimensional MFCCs consisting of 13 cepstra, delta and delta-delta coefficients are used as input.
All layers have 100 hidden units, apart from the last hidden layer, which has 39 units.
\todo{Ewald: Are we sure that these configuration details are correct (penultimate or final layer)? Herman should confirm, please. Update: I changed it to final.}
\todo[color=green]{I think it is the final layer!}
\todo[color=gray]{HK: I checked the code. The network architecture has a 39 dimensional hidden layer, which then feeds into the final output layer, which is also 39-dimensional. So if you count the very last layer as the `final layer', then this is the `penultimate layer', but this is confusing so I just changed it to `last hidden layer'. Hope that is clear to everyone?}
This layer provides the features used in the AE$_\textrm{MFCC}$ and AE$_\textrm{BNF}$ experiments.
This last hidden layer feeds into a linear output layer, producing the predicted MFCC vector.
\todo[color=gray]{Added this last sentence, which hopefully clarifies things.}

\subsection{Correspondence autoencoder features}
\label{sec:cAE}

\begin{figure}[!t]
	\includegraphics[width=0.90\linewidth]{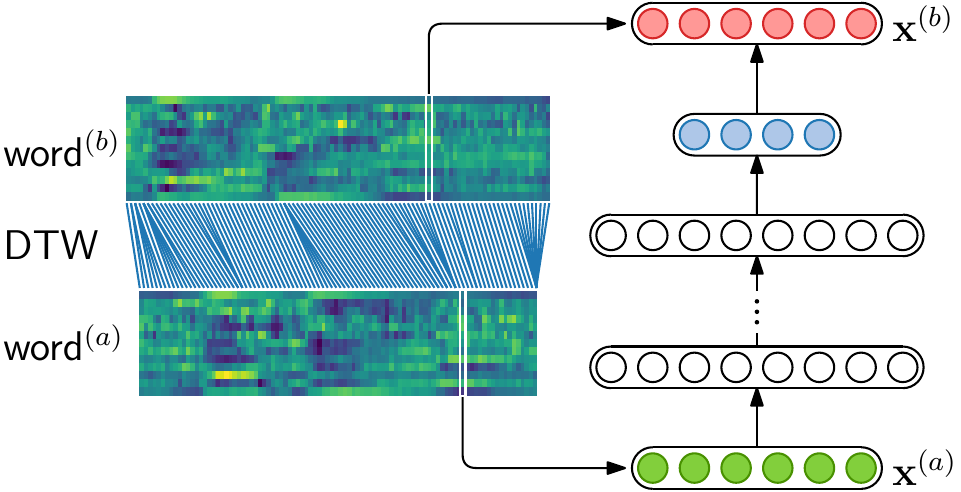}
	\caption{The correspondence autoencoder (CAE) is trained to reconstruct a frame in one word from a frame in another.}
	\label{fig:cae}
	\vspace*{-12pt}
\end{figure}

While an AE is trained using the same speech frames as input and output, a correspondence autoencoder (CAE) uses frames from different instances of the same keyword type as input and output.
Using the set of isolated keywords, we consider all possible pairs of words of the same type.
For each pair, DTW is used to find the minimum-cost frame-level alignment between the two words, as illustrated in Figure~\ref{fig:cae}.
Individual aligned frame pairs are then used as input-output pairs to the CAE.
The CAE is therefore trained on pairs of speech features $(\mathbf{x}^{(a)}, \mathbf{x}^{(b)})$, where $\mathbf{x}^{(a)}$ is a frame from one keyword, and $\mathbf{x}^{(b)}$ the corresponding aligned frame from another keyword of the same type.
Given input $\mathbf{x}^{(a)}$, the output of the network $\mathbf{y}$ is then trained to minimise the the CAE loss $||\mathbf{y} - \mathbf{x}^{(b)}||^2$, as shown in Figure~\ref{fig:cae}. 

To obtain useful features, it is essential to pretrain the CAE as a conventional AE~\cite{kamper2015unsupervised}.
Our CAE has the same structure as the AE described in Section~\ref{sec:AE} and pretraining follows the same procedure described there.
The pretrained network is then fine-tuned on the set of isolated keywords using the CAE loss described above.
Hence, the CAE takes advantage of a large amount of untranscribed data for initialisation, and then combines this with a weak form of supervision on a small amount of labelled keyword data.
Output features are extracted from the last 39-dimensional hidden layer.
\todo{Ewald: Are we sure that these configuration details are correct (penultimate or final layer)? Herman should confirm, please. Update: I changed it to final layer.}
\todo[color=green]{I think it should be the final layer and not penultimate.}
\todo[color=gray]{HK: Changed it to last hidden layer}

The intention is to use the CAE to obtain features that are insensitive to factors not common to keyword pairs, such as speaker, gender and channel, while remaining dependent on factors that are, such as the word identity.
Furthermore, the number of input-output pairs on which the CAE is fine-tuned is much larger than the total number of frames in the keyword segments themselves, because all pairwise combinations of different instances of a keyword type are considered.
For example, for the SABN dataset, the keywords contain approximately 120k frames in total, while the pairwise combinations yield approximately two million unique aligned frame pairs.
Furthermore, frame pairs are presented to the CAE in both input-output directions, thereby doubling the number of training instances to four million.

\input{results_table_english.tex}

\subsection{Bottleneck features}
\label{sec:bnf}
Multilingual bottleneck feature (BNF) extractors trained on a set of well-resourced languages have been shown to perform well in a number of studies~\cite{Menon_SLTU2018,Hermann2018,hermann+etal_submitted18,vesely2012language,vu2012investigation, cui2015multilingual, alumae2016improved,sthomas2012,chen2017multilingual, yuan2017extracting}, and can be applied directly in an almost zero-resource setting.
BNFs are obtained by training a deep neural network jointly on transcribed data from multiple languages. 
The lower layers of the network are shared among all languages.
The output layer has phone or HMM state labels as targets and may either be shared by or be separate for each language.
The layer directly preceding the output layer often has a lower dimensionality than the preceding layers, because it should capture aspects that are common to all the languages, hence, the term ``bottleneck.''

Different neural network architectures can be used to obtain BNFs.
We used the 6-layer time-delay neural networks (TDNN) trained on 10 languages from the GlobalPhone corpus described in~\cite{Hermann2018}. 
The network uses ReLU activations and batch normalisation, with a 39-dimensional bottleneck layer.
40-dimensional high resolution MFCCs appended with 100-dimensional i-vectors for speaker adaptation are used as inputs to the network.

\section{Experimental setup}
In addition to MFCCs, we use each of the neural networks described above as feature extractors, using features from the intermediate/bottleneck layers of the CAE, AE and BNF as input to our DTW-based keyword spotter.
All the neural networks take MFCCs as input.
Each takes advantage of resources in a particular way: the AE is trained on untranscribed target language data; the CAE is initialised on untranscribed data and then fine-tuned on a small amount of labelled target language data; and the BNFs use larger amounts of labelled non-target language data.
The complimentary effect of these approaches are also investigated by performing experiments in which the AE and CAE are trained with BNFs rather than MFCCs as input.
Hyperparameters for the CAE were taken directly from~\cite{kamper2015unsupervised}, i.e., no further tuning was performed on the development set, hence, it can be considered a second test~set.

Keyword spotting performance is assessed using a number of standard metrics.
The receiver operating characteristic (ROC) is obtained by plotting the false positive rate against the true positive rate as the keyword detection threshold is varied. The area under this curve (AUC) is used as a single metric across all operating points.
The equal error rate (EER) is the point at which the false positive rate equals the false negative rate, i.e.\ a lower EER indicates better system performance.
Precision at 10 ($P$@10) and precision at $N$ ($P$@$N$) are the proportion of correct keyword detections among the top 10 and top $N$ hits, respectively.

\section{Results}
\label{sec:exp_res}
The keyword spotting results for both languages are presented in Table~\ref{tab:res1}.
\todo[color=gray]{HK: I added an array stretch in both tables to just spread the results out a little bit.}
The column headings with \textsc{`mfcc'} and \textsc{`bnf'} are used to distinguish between networks trained using MFCCs and BNFs as input features.
The results for MFCC, AE$_\textrm{MFCC}$ and CAE$_\textrm{MFCC}$ features show that the CAE consistently outperforms the MFCC baseline, while the AE does not provide any benefit in this case.
The BNF and CAE$_\textrm{MFCC}$ results are comparable in the case of SABN English, while BNFs outperform CAE$_\textrm{MFCC}$ for Luganda.
Using a small amount of labelled data in a target language can therefore be just as beneficial as using large amounts of labelled data from several non-target languages for feature learning.
This may be important in situations where large out-of-domain datasets are not available.

Our best overall model on both the development and test data is the CAE$_\textrm{BNF}$.
It achieves precision values of approximately 1.7 times better than the closest competitor, while the AUC and EER are approximately 7--9\% and 4--10\% better than standard BNFs, respectively.
Compared to the baseline MFCCs, AUC and EER improve by 8--12\% when using the CAE$_\textrm{BNF}$ features.
The AE$_\textrm{BNF}$ can also achieve improvements over its MFCC counterpart, but not to the same degree as CAE$_\textrm{BNF}$.
The CAE$_\textrm{BNF}$ shows the benefits of incorporating features learned from well-resourced non-target languages with fine-tuning on a small amount of labelled target language data after pretraining on untranscribed in-domain speech.
We show this directly in an extrinsic keyword spotting task that uses features obtained from a lightly supervised neural network model.
In contrast to the work of~\cite{Hermann2018,hermann+etal_submitted18}, where discovered word pairs were used for unsupervised CAE training and the benefit of CAE training on top of BNFs were inconclusive, we obtain consistent improvements in our setting.
\todo[color=gray]{Added this last sentence; I am happy for it to be removed, but I think it would be helpful to refer back to this work again (although it is mentioned in the Intro).}


\section{Conclusion}
\label{sec:CON}

We investigated the use of different neural network features for improving ASR-free DTW-based keyword spotting in an almost zero-resource setting.
The only labelled data used were a small number of isolated keyword utterances.
Features were extracted using a multilingual bottleneck network (BNF), a stacked autoencoder (AE) and a correspondence autoencoder (CAE).
We also considered combining these, feeding the AE and CAE with BNFs instead of MFCCs.
The best performance was achieved with a CAE trained on BNFs.
This model combines the benefit of labelled data in well-resourced out-of-domain languages with a technique that can be used on extremely sparse in-domain data. 
Another interesting finding is that, in the absence of multilingual resources to train a BNF extractor, features from a CAE trained on MFCCs can yield comparable performance.
Future work includes integrating this model into our larger keyword spotting framework~\cite{Menon_Inter2018} and applying it to languages such as Somali, Rutooro and Lugbara, which are spoken in areas where the system will be deployed next.

\section{Acknowledgements}

We thank NVIDIA Corporation for donating GPU equipment used for this work, and acknowledge the support of Telkom South Africa.
HK is supported by a Google Faculty Award.
We would also like to thank Enno Hermann for assisting with bottleneck feature extraction.

\bibliographystyle{IEEEtran}

\bibliography{mybib}


\end{document}

%% file: results_table_english.tex
\begin{table*}[!t]
\caption{English and Luganda keyword spotting performance on development and test data using the different feature representations.
Subscripts are in the column headings to distinguish whether MFCCs or BNFs were used as inputs to the AE and CAE.
}
\label{tab:res1}
\centering
\renewcommand{\arraystretch}{1.2}
\begin{tabular}{lcccccccccc}
\toprule
\multirow{2}{*}{\parbox{\widthof{Metric}}{\textbf{\vspace{-1ex}Metric}}} & \multicolumn{6}{c}{\textit{{Development} (\%)}} & \multicolumn{4}{c}{\textit{{Test}} (\%)} \\
\cmidrule(lr){2-7} \cmidrule(l){8-11}
& \textbf{MFCC} & \textbf{AE}$_\textrm{\textbf{MFCC}}$ & \textbf{CAE}$_\textrm{\textbf{MFCC}}$ & \textbf{BNF} & \textbf{AE}$_\textrm{\textbf{BNF}}$ & \textbf{CAE}$_\textrm{\textbf{BNF}}$ & \textbf{MFCC} & \textbf{CAE}$_\textrm{\textbf{MFCC}}$ & \textbf{BNF} & \textbf{CAE}$_\textrm{\textbf{BNF}}$ \\
\midrule
\multicolumn{11}{c}{\bf English} \\
AUC     &                       73.32 &                       73.01 & 77.14 & 77.81 & 78.38 & \textbf{86.98} &                       74.10 & 76.86 & 76.99 & \textbf{86.39} \\
EER     &                       32.34 &                       33.51 & 28.91 & 28.72 & 28.23 & \textbf{19.24} &                       32.19 & 30.05 & 30.12 & \textbf{20.12} \\
$P$@10  &                       15.75 &                       16.50 & 25.25 & 17.00 & 17.75 & \textbf{42.25} &                       17.00 & 30.25 & 22.75 & \textbf{45.75} \\
$P$@$N$ & \leavevmode\hphantom{0}9.43 & \leavevmode\hphantom{0}9.68 & 14.66 & 13.99 & 13.64 & \textbf{30.88} & \leavevmode\hphantom{0}9.75 & 16.45 & 12.85 & \textbf{29.99} \\
\midrule
\multicolumn{11}{c}{\bf Luganda} \\
AUC     & 66.51 & 67.52 & 69.62 &                       71.24 & 72.73 & \textbf{78.09} & 69.57 & 69.74 & 73.33 & \textbf{80.59} \\
EER     & 38.68 & 38.57 & 37.20 &                       33.26 & 31.20 & \textbf{29.33} & 37.20 & 37.24 & 33.73 & \textbf{29.00} \\
$P$@10  & 11.43 & 14.29 & 28.57 &                       11.43 & 10.00 & \textbf{45.71} & 18.89 & 27.78 & 26.11 & \textbf{41.67} \\
$P$@$N$ & 10.72 & 10.13 & 13.95 & \leavevmode\hphantom{0}9.99 & 11.61 & \textbf{26.11} & 13.87 & 18.77 & 18.21 & \textbf{28.80} \\
\bottomrule
\end{tabular}
\end{table*}

%% file: main_is2019.bbl
\begin{thebibliography}{10}
\providecommand{\url}[1]{#1}
\csname url@samestyle\endcsname
\providecommand{\newblock}{\relax}
\providecommand{\bibinfo}[2]{#2}
\providecommand{\BIBentrySTDinterwordspacing}{\spaceskip=0pt\relax}
\providecommand{\BIBentryALTinterwordstretchfactor}{4}
\providecommand{\BIBentryALTinterwordspacing}{\spaceskip=\fontdimen2\font plus
\BIBentryALTinterwordstretchfactor\fontdimen3\font minus
  \fontdimen4\font\relax}
\providecommand{\BIBforeignlanguage}[2]{{%
\expandafter\ifx\csname l@#1\endcsname\relax
\typeout{** WARNING: IEEEtran.bst: No hyphenation pattern has been}%
\typeout{** loaded for the language `#1'. Using the pattern for}%
\typeout{** the default language instead.}%
\else
\language=\csname l@#1\endcsname
\fi
#2}}
\providecommand{\BIBdecl}{\relax}
\BIBdecl

\bibitem{Menon2017}
R.~Menon \emph{et~al.}, ``Radio-browsing for developmental monitoring in
  {U}ganda,'' in \emph{Proc. ICASSP}, 2017.

\bibitem{Saeb2017}
A.~Saeb \emph{et~al.}, ``Very low resource radio browsing for agile
  developmental and humanitarian monitoring,'' in \emph{Proc. Interspeech},
  2017.

\bibitem{Larson12}
M.~Larson and G.~J.~F. Jones, ``Spoken content retrieval: A survey of
  techniques and technologies,'' \emph{Found. Trends Inform. Retrieval},
  vol.~5, no. 4-5, pp. 235--422, 2012.

\bibitem{mandal2014recent}
A.~Mandal, K.~P. Kumar, and P.~Mitra, ``Recent developments in spoken term
  detection: a survey,'' \emph{Int. J. of Speech Technol.}, vol.~17, no.~2, pp.
  183--198, 2014.

\bibitem{audhkhasi+etal_icassp17}
K.~Audhkhasi, A.~Rosenberg, A.~Sethy, B.~Ramabhadran, and B.~Kingsbury,
  ``End-to-end {ASR}-free keyword search from speech,'' in \emph{Proc. ICASSP},
  2017.

\bibitem{Menon_Inter2018}
R.~Menon, H.~Kamper, J.~Quinn, and T.~Niesler, ``Fast {ASR}-free and almost
  zero-resource keyword spotting using {DTW} and {CNN}s for humanitarian
  monitoring,'' in \emph{Proc. Interspeech}, 2018.

\bibitem{Menon_SLTU2018}
------, ``{ASR}-free {CNN-DTW} keyword spotting using multilingual bottleneck
  features for almost zero-resource languages,'' in \emph{Proc. SLTU}, 2018.

\bibitem{SainathPara2015}
T.~N. Sainath and C.~Parada, ``Convolutional neural networks for
  small-footprint keyword spotting,'' in \emph{Proc. Interspeech}, 2015.

\bibitem{chen2014small}
G.~Chen, C.~Parada, and G.~Heigold, ``Small-footprint keyword spotting using
  deep neural networks.'' in \emph{Proc. ICASSP}, 2014.

\bibitem{tang2018deep}
R.~Tang and J.~Lin, ``Deep residual learning for small-footprint keyword
  spotting,'' in \emph{Proc. ICASSP}, 2018.

\bibitem{Hazen2009}
T.~J. Hazen, W.~Shen, and C.~White, ``Query-by-example spoken term detection
  using phonetic posteriorgram templates,'' in \emph{Proc. ASRU}, 2009.

\bibitem{Zhang2009}
Y.~Zhang and J.~R. Glass, ``Unsupervised spoken keyword spotting via segmental
  {DTW} on gaussian posteriorgrams,'' in \emph{Proc. ASRU}, 2009.

\bibitem{park+glass_taslp08}
A.~S. Park and J.~R. Glass, ``Unsupervised pattern discovery in speech,''
  \emph{IEEE Trans. Audio, Speech, Language Process.}, vol.~16, no.~1, pp.
  186--197, 2008.

\bibitem{jansen+vandurme_interspeech12}
A.~Jansen and B.~Van~Durme, ``Indexing raw acoustic features for scalable zero
  resource search,'' in \emph{Proc. Interspeech}, 2012.

\bibitem{vavrek2012tuke}
J.~Vavrek, M.~Pleva, and J.~Juh{\'a}r, ``Tuke mediaeval 2012: Spoken web search
  using {DTW} and unsupervised {SVM},'' in \emph{MediaEval}, 2012.

\bibitem{carlin2011rapid}
M.~A. Carlin, S.~Thomas, A.~Jansen, and H.~Hermansky, ``Rapid evaluation of
  speech representations for spoken term discovery,'' in \emph{Proc.
  Interspeech}, 2011.

\bibitem{jansen2012jhu}
A.~Jansen, B.~Van~Durme, and P.~Clark, ``The {JHU}-{HLTCOE} spoken web search
  system for {M}edia{E}val 2012.'' in \emph{MediaEval}, 2012.

\bibitem{lopez2016finding}
P.~Lopez-Otero, L.~Docio-Fernandez, and C.~Garcia-Mateo, ``Finding relevant
  features for zero-resource query-by-example search on speech,'' \emph{Speech
  Commun.}, vol.~84, pp. 24--35, 2016.

\bibitem{dunbar2017zero}
E.~Dunbar \emph{et~al.}, ``The {Zero Resource Speech Challenge} 2017,'' in
  \emph{Proc. ASRU}, 2017.

\bibitem{versteegh2016zero}
M.~Versteegh, X.~Anguera, A.~Jansen, and E.~Dupoux, ``The {Zero Resource Speech
  Challenge} 2015: Proposed approaches and results,'' in \emph{Proc. SLTU},
  2016.

\bibitem{renshaw+etal_interspeech15}
D.~Renshaw, H.~Kamper, A.~Jansen, and S.~J. Goldwater, ``A comparison of neural
  network methods for unsupervised representation learning on the {Zero
  Resource Speech Challenge},'' in \emph{Proc. Interspeech}, 2015.

\bibitem{Hermann2018}
E.~Hermann and S.~J. Goldwater, ``Multilingual bottleneck features for subword
  modeling in zero-resource languages,'' in \emph{Proc. Interspeech}, 2018.

\bibitem{hermann+etal_submitted18}
E.~Hermann, H.~Kamper, and S.~J. Goldwater, ``Multilingual and unsupervised
  subword modeling for zero resource languages,'' \emph{In Submission}, 2018.

\bibitem{vesely2012language}
K.~Vesel{\`y}, M.~Karafi{\'a}t, F.~Gr{\'e}zl, M.~Janda, and E.~Egorova, ``The
  language-independent bottleneck features,'' in \emph{Proc. SLT}, 2012.

\bibitem{vu2012investigation}
N.~T. Vu, W.~Breiter, F.~Metze, and T.~Schultz, ``An investigation on
  initialization schemes for multilayer perceptron training using multilingual
  data and their effect on asr performance,'' in \emph{Proc. Interspeech},
  2012.

\bibitem{cui2015multilingual}
J.~Cui \emph{et~al.}, ``Multilingual representations for low resource speech
  recognition and keyword search,'' in \emph{Proc. ASRU}, 2015.

\bibitem{alumae2016improved}
T.~Alum{\"a}e, S.~Tsakalidis, and R.~M. Schwartz, ``Improved multilingual
  training of stacked neural network acoustic models for low resource
  languages,'' in \emph{Proc. Interspeech}, 2016.

\bibitem{sthomas2012}
S.~Thomas, S.~Ganapathy, and H.~Hermansky, ``Multilingual mlp features for
  low-resource lvcsr systems,'' in \emph{Proc. ICASSP}, 2012.

\bibitem{chen2017multilingual}
H.~Chen, C.-C. Leung, L.~Xie, B.~Ma, and H.~Li, ``Multilingual bottle-neck
  feature learning from untranscribed speech,'' in \emph{Proc. ASRU}, 2017.

\bibitem{yuan2017extracting}
Y.~Yuan \emph{et~al.}, ``Extracting bottleneck features and word-like pairs
  from untranscribed speech for feature representation,'' in \emph{Proc. ASRU},
  2017.

\bibitem{kamper2015unsupervised}
H.~Kamper, M.~Elsner, A.~Jansen, and S.~Goldwater, ``Unsupervised neural
  network based feature extraction using weak top-down constraints,'' in
  \emph{Proc. ICASSP}, 2015.

\bibitem{Kamper2015}
H.~Kamper, F.~De~Wet, T.~Hain, and T.~Niesler, ``Capitalising on {N}orth
  {A}merican speech resources for the development of a {S}outh {A}frican
  {E}nglish large vocabulary speech recognition system,'' \emph{Comput. Speech
  Language}, vol.~28, no.~6, pp. 1255--1268, 2014.

\bibitem{gehring2013extracting}
J.~Gehring, Y.~Miao, F.~Metze, and A.~Waibel, ``Extracting deep bottleneck
  features using stacked auto-encoders,'' in \emph{Proc. ICASSP}, 2013.

\bibitem{zeiler+etal_icassp13}
M.~D. Zeiler \emph{et~al.}, ``On rectified linear units for speech
  processing,'' in \emph{Proc. ICASSP}, 2013.

\bibitem{badino2014auto}
L.~Badino, C.~Canevari, L.~Fadiga, and G.~Metta, ``An auto-encoder based
  approach to unsupervised learning of subword units,'' in \emph{Proc. ICASSP},
  2014.

\bibitem{hinton2012deep}
G.~Hinton \emph{et~al.}, ``Deep neural networks for acoustic modeling in speech
  recognition: The shared views of four research groups,'' \emph{IEEE Signal
  Process. Mag.}, vol.~29, no.~6, pp. 82--97, 2012.

\bibitem{deng2010binary}
L.~Deng, M.~L. Seltzer, D.~Yu, A.~Acero, A.~R. Mohamed, and G.~Hinton, ``Binary
  coding of speech spectrograms using a deep auto-encoder,'' in \emph{Proc.
  Interspeech}, 2010.

\bibitem{sainath2012auto}
T.~N. Sainath, B.~Kingsbury, and B.~Ramabhadran, ``Auto-encoder bottleneck
  features using deep belief networks,'' in \emph{Proc. ICASSP}, 2012.

\end{thebibliography}
